# Intermediate-Range Order in Water Ices

T.T. Fister[1,2], K.P. Nagle[1], F.D. Vila[1], G.T. Seidler[1,(*)], C. Hamner[1],

J.O. Cross[3], J.J. Rehr[1]

1. Physics Department, University of Washington, Seattle, Washington 98105
2. Materials Science Division, Argonne National Laboratory, Argonne, IL 60637
3. Advanced Photon Source, Argonne National Laboratory, Argonne, IL 60637

We report measurements of the non-resonant inelastic x-ray scattering (NRIXS) from the O 1*s* orbitals in ice Ih, and also report calculations of the corresponding spectra for ice Ih and several other phases of water ice. We find that the intermediate-energy fine structure may be calculated well using an *ab initio* real-space full multiple scattering approach, and that it provides a strong fingerprint of the intermediate-range order for some ice phases. These results have important consequences for future NRIXS measurements of high-pressure phases of ice and also may call into question the assumption that the wavefunctions for final states within a few eV of the absorption edge are strongly localized.

I. INTRODUCTION

The bulk and surface structure of the various phases of water ice is of considerable contemporary importance. In atmospheric sciences, the surface structure of ice and the occurrence of surface premelting may play a major role in the chemistry of the troposphere.[1] In geosciences and planetary sciences, the many phases of ice hold interest for understanding the composition of comets,[2] of planetary cores in the outer reaches of the solar system,[3] and of subsurface and atmospheric regions of Mars.[4] Finally, in astrophysics, various phases of water ice are present in interstellar dust.[5]

In recent years, the use of synchrotron radiation techniques together with improved modeling and theoretical treatment has reopened vigorous debate about the local structure of water[6-10] and has also revealed new aspects of the surface structure and properties of water ices.[11,12] This ensemble of results includes new measurements and interpretation of x-ray absorption fine-structure (XAS),[7,12-15] and the valence[9,16] (Compton scattering) and core[8,10,17-19] (x-ray Raman scattering, henceforth XRS) contributions to nonresonant inelastic x-ray scattering (NRIXS) of hard x-rays. The two NRIXS techniques show promise for studies of the many high-pressure phases of water ice because of the compatibility of the incident hard x-rays ($\hbar\omega \geq 10\,\text{keV}$) with high pressure chambers.[18,20,21]

Here, we present improved measurements and calculations of XRS from water ice, with an emphasis on the importance of the intermediate-energy fine structure for fingerprinting subtle structural changes in higher coordination shells. On the experimental front, we present new XRS measurements of the O $K$-edge in ice Ih with greatly improved statistics compared to prior measurements. This enables a careful discussion of previously unobserved intermediate-energy XRS fine structure while also establishing that the momentum-transfer (*q*) dependence of XRS from water ices can be ignored when away from the absorption edge. This second result has significant practical value for future XRS studies of high-pressure ice phases, as it endorses measurement at high scattering angles (*i.e.*, high *q*) where the overall count rate is typically more than an order of magnitude larger than at low scattering angles. We also present new theoretical results for the XRS or XAS fine structure using *ab initio* real-space full multiple scattering calculations on large clusters of several different water ice phases. In contrast to prior RSFMS calculations on small clusters or to prior semi-empirical molecular-orbital based calculations which are restricted to a small energy range near the edge, we obtain qualitative agreement between calculation and experiment both in the very near-edge and intermediate-energy regimes. These calculations demonstrate the sensitivity of at least the intermediate-energy fine structure in water ices to relatively high coordination shells, i.e., the intermediate-range order in the various phases.

**II. THEORY**

For a powder or amorphous sample, the double-differential cross-section for an NRIXS measurement is

$$\frac{d^2\sigma_{XRS}}{d\Omega d\omega} \propto S(q,\omega) = \sum_f \left| \langle f | e^{iqr} | i \rangle \right|^2 \delta(E_f - E_i + \hbar\omega), \qquad (1)$$

where $S(q,\omega)$ is the dynamic structure factor, the *i* and *f* indices refer to the initial and final states, and $\hbar\omega$ is the energy loss in the scattering event. For sufficiently small *q*, $S(q,\omega)$ becomes proportional to the usual x-ray absorption coefficient such as is measured by inherently dipole-limited x-ray absorption spectroscopy (XAS) or in typical electron energy loss spectroscopy (EELS) studies.[21-24] This connection between XAS and low-$q$ XRS has led to numerous recent studies where the relatively large penetration length of the incident hard x-rays used in XRS has provided a bulk-sensitive and pressure-chamber compatible alternative to soft x-ray XAS.[8,19,25] At higher $q$, additional (*i.e.* non-dipole) angular momentum selection rules become important, allowing for a more detailed characterization of the symmetry of the final states[24,26,27] in a way unavailable to XAS but which has been exploited in gas-phase inner-shell EELS studies[22] at higher momentum transfers. In some cases, $q$-dependent XRS measurements may be inverted for a purely experimental determination of the symmetry-projected final density of states.[28]

Given the close connection between XRS and XAS, some of the theoretical treatments commonly used in XAS have now been extended to XRS[29,30] In the present study, we are concerned with the near-edge structure extended to perhaps 50 eV past the O *K*-edge for ordered and weakly disordered phases of water ice. We address these features with real-space full multiple scattering (RSFMS) approach using a $q$-dependent version[30] of the FEFF software package. This code computes $S(q,\omega)$ with a single particle Greens function using self-consistent muffin-tin potentials.[30,31] The real-space treatment in FEFF does not require periodicity and can accommodate large atomic clusters with or without disorder. The ability to treat a large disordered cluster (~200 molecules) is crucial, as our calculations find sensitivity of the XRS (or XAS) near-edge structure to subtle changes in high coordination shells, as we discuss below.

## III. EXPERIMENT

All XRS measurements were taken at the Advanced Photon Source, sector 20ID-PNC/XOR using the lower energy resolution inelastic x-ray scattering (LERIX) user facility.[32] A sample of HPLC water was sealed into a 1-cm thick Al chamber with 25 μm-thick polyimide windows and was slowly cooled in vacuum by a nitrogen-flow cryostat to ~130 K. Such preparation is well-known to result in ice Ih. There were no systematic changes between spectra collected over several hours. The energy resolution was 1.35 eV.

The sample surface was tilted back ~30 deg from the vertical, so as to optimize the high-angle scattering. XRS spectra were simultaneously collected at three $q$ ranging from 7.7-9.3 Å$^{-1}$. These values correspond to $qa = 0.8 – 1.0$, where $a$ is the average radius of the O 1$s$ initial state. For this experimental geometry, we expect less than 1% percent background contribution from the polyimide window on the sample cell. The rationale for emphasizing high-$q$ (high scattering angle) spectra was pragmatic: the overall XRS cross-section increases nearly as $q^2$ over the available $q$ range.[30] For each $q$, we $f$-sum normalize the data and remove the valence Compton background by fitting to a Gaussian.[27] Integrating over $q$, we then obtain 1.5 x 10$^5$ counts above the valence Compton background at the O $K$-edge.

## III. RESULTS AND DISCUSSION

In Figure 1, we show our new XRS measurements (top curve, labeled 'XRS') and the same data after improving the effective energy resolution to 0.9 eV by use of the Richardson-Lucy iterative deconvolution algorithm[33] (middle curve, 'XRS+RL'). Despite using momentum transfers outside the nominal dipole-scattering limit (i.e., $qa \ll 1$) the observed spectra have good agreement with the previous dipole-limited XAS results (Fig.1, bottom curve) of Zubavichus, *et al.*[14] This agreement is likely due to the relatively low symmetry of the O local environment: the local electronic structure must be a strong admixture of *s*, *p*, *d*, and possibly higher orbital angular momentum contributions. Recent work on the extended fine structure measured in XRS from water came to similar conclusions.[19]

Briefly, the fine structure in XAS or in XRS with increasing photoelectron kinetic energy (KE) is due to interference between the outgoing photoelectron wavefunction and its many reflections off of neighboring atoms, with phase shifts mediated by the details of the atomic and interatomic potentials.[34] At high photoelectron KE (i.e., in the limit of extended XAS), a small number of paths typically dominate and the fine details of the potentials (such as bonding anisotropy) are largely irrelevant. For ice Ih, the extended fine structure in XRS was measured first by Bowron, *et al.*,[35] and most recently by Bergmann, *et al.*,[19] with considerably improved statistics.

On the other hand, at very low photoelectron KE, i.e., within ~10 eV of the edge, the photoelectron scattering is isotropic, intrinsic losses are low, and the fine details of the potentials may strongly influence the phase shifts determining the exact interference pattern. This is especially true if the final states are very localized, such as single-atomic-like states or antibonding molecular orbitals.[36] This energy loss regime has been the subject of intensive recent interest and contentious debate for liquid water. [6-10]

Finally, for more intermediate photoelectron KE (~10 – 50 eV), intrinsic losses are still small, the generic (but not exact) details of the potentials are important, the final states are now definitely spatially extended, and the photoelectron scattering is still relatively isotropic. An RSFMS approach[31] using a large atomic cluster is therefore physically appropriate. Due to the isotropic scattering and low losses, features in this regime are often more sensitive to higher coordination shells than for either lower or higher photoelectron KE. For example, for ice Ih we expect that the low-energy photoelectron inelastic mean free path is greater than 10 Å. In this energy loss regime XAS and XRS may therefore be sensitive to intermediate-range order, rather than just short-range order.

We first consider ice Ih using a model whose coordinates were generated using an interaction potential based on a single-center multipole expansion, which incorporates hydrogen disorder via Pauling's criteria before optimizing the coordinates to match the experimental crystal lattice energy and lattice parameters.[37] Convergence of the self-consistent calculation of the atomic potentials requires a 4 Å cluster with $l_{max} = 2$ (i.e., including *s*-, *p*-, and *d*-type unoccupied states in the self-consistent calculation). The calculations are not strongly affected by the number of unique potentials and thus select the simplest configuration: one unique potential for the absorber, one for the hydrogens, and one for the remaining oxygens. In the near-edge region, best agreement between experiment and theory occurs when using the default Hedin-Lundquist exchange-correlation potential, but omitting the core-hole from the calculation.

With the potentials thus specified, we proceeded to vary the hydrogen muffin tin radius while keeping the oxygen potential fixed as shown in Figure 2. This test is essential for hydrogen atoms whose charge density is highly anisotropic. The simulations in the figure were performed for $q = 8.0$ Å$^{-1}$, but the results beyond 537 eV are essentially unchanged at smaller $q$.[38] For subsequent calculations, we choose the potential with a muffin-tin radius of 0.47 Å for its general agreement with the labeled features.

We next turn to the problem of convergence of the RSFMS calculation. The spatial extent of the photoelectron wavefunction for the various spectral features may be investigated by varying the size of the cluster used in the calculation. In Figure 3, we show calculations for ice Ih for $q = 8.0$ Å$^{-1}$ for gradually increasing cluster size; again, the results are essentially unchanged at smaller $q$ so that the same calculations may also be compared to XAS spectra. A 7 Å radius cluster consisting of 46 molecules (approximately the first 5 coordination shells) is necessary for convergence of the RSFMS calculation in the intermediate-energy regime. This result is consistent with the hypothesis of Parent, *et al*,[15] and Zubavichus, *et al*,[14] who proposed that the intermediate-energy region would be sensitive to intermediate-range order in ice, but who were unsuccessful in reproducing this fine structure using RSFMS calculations. This is apparently due to the small size of the clusters used in those calculations.

The steady evolution of the very near-edge region (i.e., within 10 eV of the absorption edge) with increasing cluster size deserves comment. The FEFF code has two limitations with regard to calculations in this region, especially for molecular solids. First, the use of spherical muffin-tin potentials may be too crude an approximation when there is strong anisotropy in the real local potentials due to chemical bonding. Second, and more importantly, this approximation can be inadequate to treat low-energy atomic-like or molecular-orbital-like bound states (resonances) if they are present. However, the qualitative agreement between the calculated spectra in the first 10 eV and the experimental results suggests that key properties of the final states in this regime are reasonably well preserved in the present calculation. This result runs contrary to the assumptions underlying the prior[7] semi-empirical, molecular-orbital based calculations used in the analysis of the first 10 eV for XAS and XRS of liquid water and ice Ih. It would be valuable to extend the present RSFMS calculations from muffin-tin potentials to a full-potential calculation to determine if this result is fortuitous or if it instead indicates that the very near-edge structure is indeed a measure of intermediate range order rather than only a measure of the local bonding configuration.

With convergence of the calculations under control, we now present our main computational results in Figure 4. From top to bottom in the figure, we show the calculated XRS spectra for the O *K*-edge for proton- and oxygen-ordered ice XI, ice Ih with oxygen disorder (via structural optimization) and proton disorder (via Pauling's criterion), ice Ih with only proton disorder, ice Ic with only proton disorder, and proton- and oxygen-ordered ice II. For systems with disorder, the displayed spectra are the result of averaging calculations at 10 different central O sites for each structure.

The spectra vary the most in the first 10 eV (i.e. for features 'A' and 'B') where the dependence on the proton disorder is most pronounced. The intermediate-energy multiple scattering features (now labeled 'C', 'D' and 'E') shift and broaden slightly in the oxygen disordered ice Ih case and agree better with the measured results. None of the calculations were able to reproduce the dip in the overall spectral shape that occurs in the XRS and XAS data at 548 eV.

There are two important comparisons to be made among the calculated spectra in this figure. First, in comparing the computed spectra for ice Ih, structurally optimized (i.e. oxygen disordered) ice Ih, and ice XI using experimental lattice parameters,[39] we found that oxygen disorder had the largest impact on the intermediate-energy fine structure (features 'C', 'D', and 'E' in Fig. 4). Of the three cases, the computed oxygen-disordered Ice Ih spectrum best matches the position and the amplitude of the experimental fine structure. Second, by comparing ice Ic and ice Ih, which have identical local structures for the first two oxygen coordination shells, we find that the relative amplitude of features 'C' and 'D' could be a useful differentiating fingerprint. We also computed the spectra for ice II, a high pressure phase of ice which is different from ice Ih even at the first coordination shell. Clearly, the calculated ice II spectrum is significantly different from ice Ih at intermediate energies, again speaking to the value of the intermediate-energy fine structure for structural determination. Discrepancies in the first 10 eV of the ice II calculation with respect to the XRS results of Cai, *et al*,[18] may benefit from further optimization with the current RSFMS theory (i.e. different exchange-correlation parameters, energy shifts, and Fermi level cutoffs) and should be reevaluated in future calculations based on more realistic potentials.

**CONCLUSIONS**

In conclusion, we have used new measurements and calculations of nonresonant x-ray Raman scattering to investigate the sensitivity of the local electronic structure to the intermediate range order in different phases of water ice. We find that the intermediate-energy fine structure (*i.e.*, 10-50 eV past the edge), which has previously been ignored in XRS studies, may be reliably calculated by real-space full multiple scattering (RSFMS) methods and that it shows significant fingerprinting for the intermediate range order, i.e., crystalline structure past the first few coordination shells. Both the theoretical and the experimental results also endorse the experimental convenience of XRS measurement at high momentum transfers, which should simplify future XRS measurements in the intermediate energy regime for high-pressure phases of water ice. Finally, the RSFMS calculations show surprisingly good agreement in the very near-edge region of the spectrum for ice Ih, which may be fortuitous or which may indicate that even this energy range involves significantly delocalized final states. Further calculations using more realistic potentials will be necessary to resolve this issue.

**ACKNOWLEDGEMENTS**

This research was supported by DOE, Basic Energy Science, Office of Science, Contract Nos. DE-FGE03-97ER45628 and W-31-109-ENG-38, ONR Grant No. N00014-05-1-0843, Grant DE-FG03-97ER5623, NIH NCRR BTP Grant RR-01209 and the Summer Research Institute Program at the Pacific Northwest National Lab. The operation of Sector 20 PNC-CAT/XOR is supported by DOE Basic Energy Science, Office of Science, Contract No. DE-FG03-97ER45629, the University of Washington, and grants from the Natural Sciences and Engineering Research Council of Canada. Use of the Advanced Photon Source was supported by the U.S. Department of Energy, Basic Energy Sciences, Office of Science, under Contract W-31-109-Eng-38. We thank Aleksi Soininen, Ed Stern, Josh Kas, and Micah Prange for stimulating discussions.

**FIGURE CAPTIONS**

**Fig. 1**: Top: The XRS near edge structure of ice Ih (present study); Middle: Present data, after partial removal of instrumental broadening (see text for details); Bottom: XAS results from Zubavichus, *et al*.[14]

**Fig. 2**: Calculated XRS spectra for ice Ih at $q = 8$ Å$^{-1}$ as a function of the muffin-tin radius at the hydrogen sites. For subsequent calculations, we chose a muffin-tin radius of 0.47 Å due to its agreement in the placement of features 'C' and 'D' and the overall shape of the spectrum in the first 10 eV.

**Fig. 3**: The upper sequence of displaced curves show the dependence on the cluster size for RSFMS calculations of the O *K*-edge XRS spectrum for ice Ih at $q = 8$ Å$^{-1}$. The curves are labeled by the cluster radius and the number of molecules that were used in the calculation. By means of reference, the bottom-most two curves are the measured XRS spectrum and the same data after partial removal of instrumental broadening.

**Fig. 4**: The sequence of curves shows the converged, large-cluster RSFMS calculations for several phases of water ice. See the text for discussion.

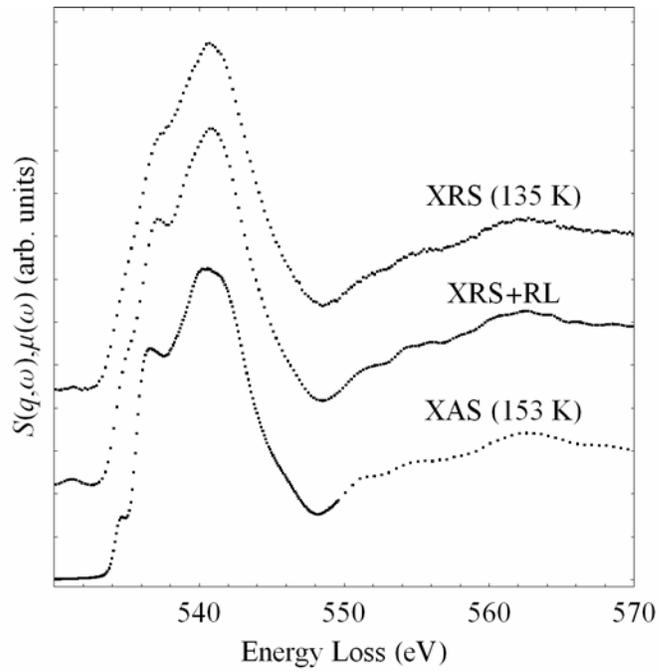

**FIGURE 1**: T.T. Fister, et al, "Intermediate Range Order in Water Ices", submitted Physical Review B (2007).

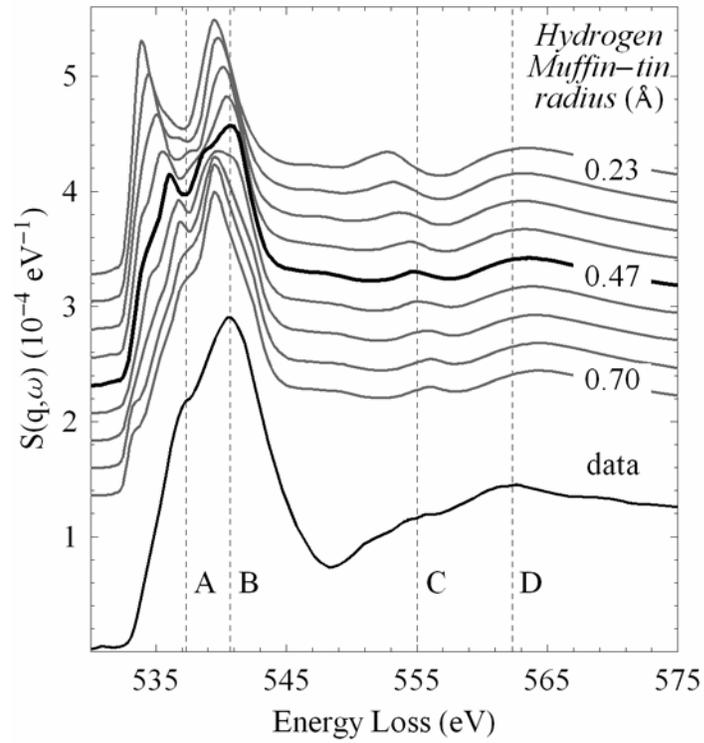

**FIGURE 2**: T.T. Fister, et al, "Intermediate Range Order in Water Ices", submitted Physical Review B (2007).

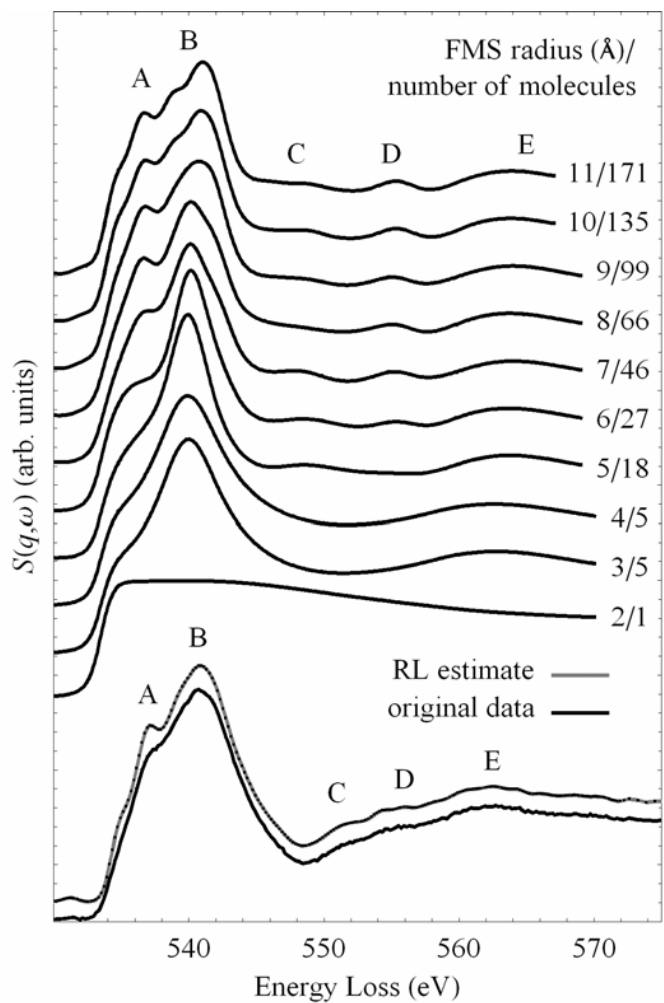

**FIGURE 3**: T.T. Fister, et al, "Intermediate Range Order in Water Ices", submitted Physical Review B (2007).

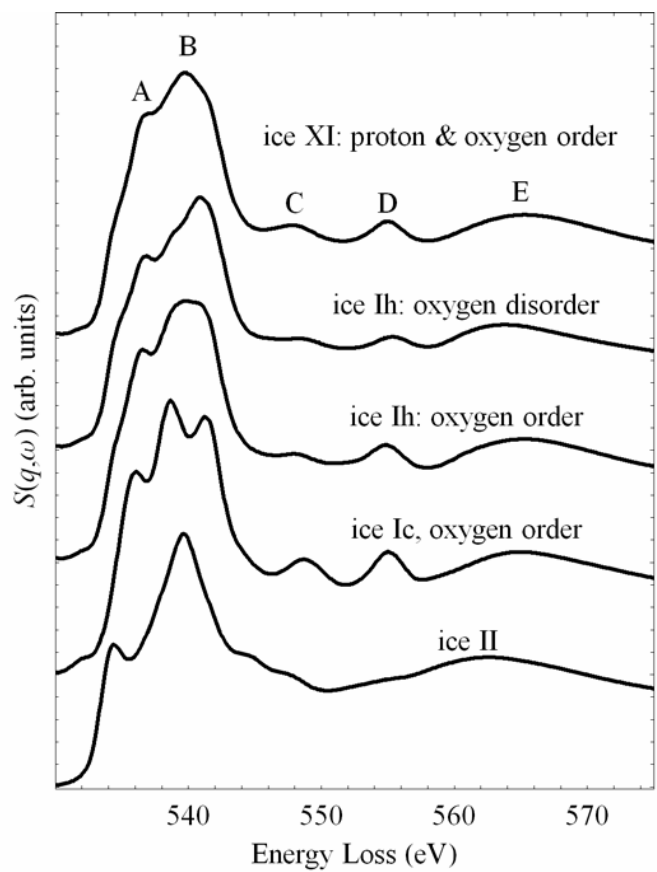

**FIGURE 4**: T.T. Fister, et al, "Intermediate Range Order in Water Ices", submitted Physical Review B (2007).

# REFERENCES

[1] J. G. Dash, H. Y. Fu, and J. S. Wettlaufer, Reports on Progress in Physics **58** (1), 115 (1995); J. E. Shilling, M. A. Tolbert, O. B. Toon, E. J. Jensen, B. J. Murray, and A. K. Bertram, Geophysical Research Letters **33** (17), L17801 (2006).

[2] J. Klinger, Icarus **47** (3), 320 (1981).

[3] P. Jenniskens and D. F. Blake, Science **265** (5173), 753 (1994).

[4] W. V. Boynton, W. C. Feldman, S. W. Squyres, T. H. Prettyman, J. Bruckner, L. G. Evans, R. C. Reedy, R. Starr, J. R. Arnold, D. M. Drake, P. A. J. Englert, A. E. Metzger, I. Mitrofanov, J. I. Trombka, C. d'Uston, H. Wanke, O. Gasnault, D. K. Hamara, D. M. Janes, R. L. Marcialis, S. Maurice, I. Mikheeva, G. J. Taylor, R. Tokar, and C. Shinohara, Science **297** (5578), 81 (2002).

[5] J. B. Pollack, D. Hollenbach, S. Beckwith, D. P. Simonelli, T. Roush, and W. Fong, Astrophysical Journal **421** (2), 615 (1994).

[6] R. Bukowski, K. Szalewicz, G. C. Groenenboom, and A. van der Avoird, Science **315** (5816), 1249 (2007); A. Nilsson, P. Wernet, D. Nordlund, U. Bergmann, M. Cavalleri, M. Odelius, H. Ogasawara, L. A. Naslund, T. K. Hirsch, P. Glatzel, and L. G. M. Pettersson, Science **308** (5723), 793A (2005); J. D. Smith, C. D. Cappa, B. M. Messer, W. S. Drisdell, R. C. Cohen, and R. J. Saykally, Journal of Physical Chemistry B **110** (40), 20038 (2006); J. D. Smith, C. D. Cappa, K. R. Wilson, B. M. Messer, R. C. Cohen, and R. J. Saykally, Science **306** (5697), 851 (2004); P. Wernet, D. Nordlund, U. Bergmann, M. Cavalleri, M. Odelius, H. Ogasawara, L. A. Naslund, T. K. Hirsch, L. Ojamae, P. Glatzel, L. G. M. Pettersson, and A. Nilsson, Science **304** (5673), 995 (2004); M. Leetmaa, M. Ljungberg, H. Ogasawara, M. Odelius, L. A. Naslund, A. Nilsson, and L. G. M. Pettersson, Journal of Chemical Physics **125** (24), 244510 (2006); R. L. C. Wang, H. J. Kreuzer, and M. Grunze, Physical Chemistry Chemical Physics **8** (41), 4744 (2006); T. Head-Gordon and M. E. Johnson, Proceedings of the National Academy of Sciences of the United States of America **103** (21), 7973 (2006); D. Prendergast and G. Galli, Physical Review Letters **96** (21), 215502 (2006); M. V. Fernandez-Serra and E. Artacho, Physical Review Letters **96** (1), 016404 (2006); L. A. Naslund, J. Luning, Y. Ufuktepe, H. Ogasawara, P. Wernet, U. Bergmann, L. G. M. Pettersson, and A. Nilsson, Journal of Physical Chemistry B **109** (28), 13835 (2005).

[7] Ph. Wernet, D. Nordlund, U. Bergmann, M. Cavalleri, M. Odelius, H. Ogasawara, L. A. Naslund, T. K. Hirsch, L. Ojamae, P. Glatzel, L. G. M. Pettersson, and A. Nilsson, Science **304**, 5 (2004).

[8] U. Bergmann, D. Nordlund, P. Wernet, M. Odelius, L. G. M. Pettersson, and A. Nilsson, Physical Review B **76** (2), 024202 (2007).

[9] K. Nygard, M. Hakala, S. Manninen, A. Andrejczuk, M. Itou, Y. Sakurai, L. G. M. Pettersson, and K. Hamalainen, Physical Review E **74** (3), 031503 (2006).

[10] A. K. Soper, Journal of Physics-Condensed Matter **17** (45), S3273 (2005).

[11] H. Bluhm, D. F. Ogletree, C. S. Fadley, Z. Hussain, and N. Salmeron, Journal of Physics-Condensed Matter **14** (8), L227 (2002).

[12] D. Nordlund, H. Ogasawara, Ph. Wernet, M. Nyberg, M. Odelius, L.G.M. Pettersson, and A. Nilsson, Chemical Physics Letters **395**, 161 (2004).

13  C. Laffon, S. Lacombe, F. Bournel, and P. Parent, Journal of Chemical Physics **125**, 204714 (2006); Y. Zubavichus, Y.J. Yang, M. Zharnikov, O. Fuchs, T. Schmidt, C. Heske, E. Umbach, G. Tzvetkov, F.P. Netzer, and M. Grunze, ChemPhysChem **5** (4), 509 (2004).

14  Y. Zubavichus, M. Zharnikov, Y.J. Yang, O. Fuchs, E. Umbach, C. Heske, and M. Grunze, Langmuir **22**, 7241 (2006).

15  P. Parent, C. Laffon, C. Mangeney, F. Bournel, and M. Tronc, Journal of Chemical Physics **117** (23), 10842 (2002).

16  M. Hakala, K. Nygard, S. Manninen, S. Huotari, T. Buslaps, A. Nilsson, L. G. M. Pettersson, and K. Hamalainen, Journal of Chemical Physics **125** (8), 084504 (2006); E. D. Isaacs, A. Shukla, P. M. Platzman, D. R. Hamann, B. Barbiellini, and C. A. Tulk, Physical Review Letters **82** (3), 600 (1999); T. K. Ghanty, V. N. Staroverov, P. R. Koren, and E. R. Davidson, Journal of the American Chemical Society **122** (6), 1210 (2000).

17  U. Bergmann, P. Wernet, P. Glatzel, M. Cavalleri, L. G. M. Pettersson, A. Nilsson, and S. P. Cramer, Physical Review B **66** (9), 092107 (2002); K. Nygard, M. Hakala, T. Pylkkanen, S. Manninen, T. Buslaps, M. Itou, A. Andrejczuk, Y. Sakurai, M. Odelius, and K. Hamalainen, Journal of Chemical Physics **126** (15), 154508 (2007).

18  Y. Q. Cai, H. K. Mao, P. C. Chow, J. S. Tse, Y. Ma, S. Patchkovskii, J. F. Shu, V. Struzhkin, R. J. Hemley, H. Ishii, C. C. Chen, I. Jarrige, C. T. Chen, S. R. Shieh, E. P. Huang, and C. C. Kao, Physical Review Letters **94** (2), 025502 (2005).

19  U. Bergmann, A. DiCicco, P. Wernet, E. Principi, P. Glatzel, and A. Nilsson, Journal of Chemical Physics **127**, 174504 (2007).

20  W. L. Mao, H. K. Mao, Y. Meng, P. J. Eng, M. Y. Hu, P. Chow, Y. Q. Cai, J. F. Shu, and R. J. Hemley, Science **314** (5799), 636 (2006).

21  U. Bergmann, P. Glatzel, and S. P. Cramer, Microchemical Journal **71** (2-3), 221 (2002); M. Krisch and F. Sette, Surface Review and Letters **9** (2), 969 (2002).

22  A. P. Hitchcock, Journal of Electron Spectroscopy and Related Phenomena **112** (1-3), 9 (2000); A. P. Hitchcock, I. G. Eustatiu, J. T. Francis, and C. C. Turci, Journal of Electron Spectroscopy and Related Phenomena **88**, 77 (1998); K. T. Leung, Journal of Electron Spectroscopy and Related Phenomena **100**, 237 (1999).

23  Y. Mizuno and Y. Ohmura, Journal of the Physical Society of Japan **22** (2), 445 (1967); H. Nagasawa, S. Mourikis, and W. Schulke, Journal of the Physical Society of Japan **58** (2), 710 (1989); T. Suzuki, Journal of the Physical Society of Japan **22** (5), 1139 (1967).

24  W. Schulke, Journal of Physics-Condensed Matter **13** (34), 7557 (2001).

25  H. Sternemann, C. Sternemann, J. S. Tse, S. Desgreniers, Y. Q. Cai, G. Vanko, N. Hiraoka, A. Schacht, J. A. Soininen, and M. Tolan, Physical Review B **75** (24), 245102 (2007); M. Balasubramanian, C. S. Johnson, J. O. Cross, G. T. Seidler, T. T. Fister, E. A. Stern, C. Hamner, and S. O. Mariager, Applied Physics Letters **91** (3), 031904 (2007); U. Bergmann, O. C. Mullins, and S. P. Cramer, Analytical Chemistry **72** (11), 2609 (2000); S. K. Lee, P. J. Eng, H. K. Mao, Y. Meng, and J. Shu, Physical Review Letters **98** (10), 105502 (2007); S. K. Lee, P. J. Eng, H. K. Mao, Y. Meng, M. Newville, M. Y. Hu, and J. F. Shu, Nature Materials **4** (11), 851 (2005); W. L. Mao, H. K. Mao, P. J. Eng, T. P. Trainor, M. Newville, C. C. Kao, D. L. Heinz, J. F. Shu, Y. Meng, and R. J. Hemley, Science **302** (5644), 425 (2003); Y. Meng, H. K. Mao, P. J. Eng, T. P. Trainor, M. Newville, M. Y. Hu, C. C. Kao, J. F. Shu, D. Hausermann, and R. J. Hemley, Nature Materials **3** (2), 111 (2004); J. F. Lin, H. Fukui, D. Prendergast, T. Okuchi, Y. Q. Cai, N. Hiraoka, C. S. Yoo, A. Trave, P. Eng, M. Y. Hu, and P. Chow, Physical Review B **75** (1), 012201 (2007).

[26] C. Sternemann, J. A. Soininen, M. Volmer, A. Hohl, G. Vanko, S. Streit, and M. Tolan, Journal of Physics and Chemistry of Solids **66** (12), 2277 (2005); Y. J. Feng, G. T. Seidler, J. O. Cross, A. T. Macrander, and J. J. Rehr, Physical Review B **69** (12), 125402 (2004); K. Hamalainen, S. Galambosi, J. A. Soininen, E. L. Shirley, J. P. Rueff, and A. Shukla, Physical Review B **65** (15), 155111 (2002); M. H. Krisch, F. Sette, C. Masciovecchio, and R. Verbeni, Physical Review Letters **78** (14), 2843 (1997); C. Sternemann, J. A. Soininen, S. Huotari, G. Vanko, M. Volmer, R. A. Secco, J. S. Tse, and M. Tolan, Physical Review B **72** (3), 035104 (2005); H. Sternemann, J. A. Soininen, C. Sternemann, K. Hamalainen, and M. Tolan, Physical Review B **75** (7), 075118 (2007); C. Sternemann, M. Volmer, J. A. Soininen, H. Nagasawa, M. Paulus, H. Enkisch, G. Schmidt, M. Tolan, and W. Schulke, Physical Review B **68** (3), 035111 (2003); A. Mattila, J. A. Soininen, S. Galambosi, S. Huotari, G. Vanko, N. D. Zhigadlo, J. Karpinski, and K. Hamalainen, Physical Review Letters **94** (24), 247003 (2005).

[27] T. T. Fister, G. T. Seidler, C. Hamner, J. O. Cross, J. A. Soininen, and J. J. Rehr, Physical Review B **74**, 214117 (2006).

[28] S. Galambosi, M. Knaapila, J. A. Soininen, K. Nygard, S. Huotari, F. Galbrecht, U. Scherf, A. P. Monkman, and K. Hamalainen, Macromolecules **39** (26), 9261 (2006); J. A. Soininen, A. Mattila, J. J. Rehr, S. Galambosi, and K. Hamalainen, Journal of Physics-Condensed Matter **18** (31), 7327 (2006); T. T. Fister, F. D. Vila, G. T. Seidler, L. Svec, J.C. Linehan, and J. O. Cross, Journal of the American Chemical Society **accepted** (2007).

[29] A. Sakko, M. Hakala, J. A. Soininen, and K. Hamalainen, Physical Review B **accepted** (2007); J. A. Soininen, K. Hamalainen, W. A. Caliebe, C. C. Kao, and E. L. Shirley, Journal of Physics-Condensed Matter **13** (35), 8039 (2001); J. A. Soininen and E. L. Shirley, Physical Review B **64** (16) (2001).

[30] J. A. Soininen, A. L. Ankudinov, and J. J. Rehr, Physical Review B (Condensed Matter and Materials Physics) **72** (4), 45136 (2005).

[31] A.L. Ankudinov, B. Ravel, J.J. Rehr, and S.D. Conradson, Physical Review B **58** (12), 7565 (1998).

[32] T. T. Fister, G. T. Seidler, L. Wharton, A. R. Battle, T. B. Ellis, J. O. Cross, A. T. Macrander, W. T. Elam, T. A. Tyson, and Q. Qian, Review of Scientific Instruments **77** (6), 063901 (2006).

[33] T. T. Fister, G. T. Seidler, J. J. Rehr, J. J. Kas, W. T. Elam, J. O. Cross, and K. P. Nagle, Physical Review B (submitted) (2007).

[34] J. J. Rehr and R. C. Albers, Reviews of Modern Physics **72** (3), 621 (2001).

[35] D. T. Bowron, M. H. Krisch, A. C. Barnes, J. L. Finney, A. Kaprolat, and M. Lorenzen, Physical Review B **62** (14), R9223 (2000).

[36] J. Stohr, *NEXAFS Spectroscopy*. (Springer, Berlin, 1992).

[37] E. R. Batista, S. S. Xantheas, and H. Jonsson, Journal of Chemical Physics **109** (11), 4546 (1998); F.D. Vila, R. Batista, and H. Jonsson, unpublished (2005).
[38] T.T. Fister, University of Washington, 2007.
[39] C. M. B. Line and R. W. Whitworth, Journal of Chemical Physics **104** (24), 10008 (1996).